\documentclass[
nofootinbib,
prfluids,aps, reprint, superscriptaddress, amsmath,amssymb,
twocolumn,
]{revtex4-2}

\usepackage{graphicx}
\usepackage{dcolumn}
\usepackage{bm}
\usepackage{hyperref}

\usepackage{physics}
\usepackage{siunitx}

\usepackage{xcolor}

\begin{document}

\preprint{APS/123-QED}

\title{Brownian Dynamics Simulations of Inclusions in an Active Fluid Bath}
\author{Lijie Ding}
\email{dingl1@ornl.gov}
\affiliation{Neutron Scattering Division, Oak Ridge National Laboratory, Oak Ridge, TN 37831, USA}
\author{Robert A. Pelcovits}
\affiliation{Department of Physics, Brown University, Providence, RI 02912, USA.}
\affiliation{Brown Theoretical Physics Center, Brown University, Providence, RI 02912, USA.}
\author{Thomas R. Powers}
\affiliation{Department of Physics, Brown University, Providence, RI 02912, USA.}
\affiliation{Brown Theoretical Physics Center, Brown University, Providence, RI 02912, USA.}
\affiliation{School of Engineering, Brown University, Providence, RI 02912, USA.}
\affiliation{Center for Fluid Mechanics, Brown University, Providence, RI 02912, USA.}

\date{\today}

\begin{abstract}
We carry out two-dimensional Brownian dynamics simulations of the behavior of rigid inclusion particles immersed in an active fluid bath. The active fluid is modeled as a collection of self-propelled circular disks interacting via a soft repulsive potential and a nematic alignment interaction. The fluid is characterized by its nematic order, polar order and orientational correlation length. The active fluid bath transitions from the isotropic to  the nematic phase with increasing number density, increasing nematic interaction strength or increasing P\'eclet number. The inclusion particles are modeled as rigid assemblies of passive circular disks. Four types of inclusions are considered: a rod-like $I$ shape, a boomerang-like $L$ shape, and stair-like shapes $Z$ and $Z^*$, with opposite handedness. When inclusions are introduced into the active fluid bath, their diffusion is significantly enhanced by the force and torque exerted by the active fluid particles and the chiral inclusion particles exhibit constant rotational drift. These diffusion and rotation enhancements increase as the swimming speed of the active fluid particles increases. The translational motion of the inclusion particles also couples with their orientational motion, and the correlation is modulated by the active fluid particles' swimming speed. This work paves the way for future simulations of inclusions in active fluid baths and suggests potential avenues for controlling transport properties in active materials.

\end{abstract}

\maketitle

\section{Introduction}

Active fluids \cite{saintillan2018rheology, marchetti2013hydrodynamics,ramaswamy2010mechanics} are non-equilibrium systems in which individual particles continuously convert stored or ambient energy into motion, resulting in rich collective behavior. From bacterial suspensions \cite{aranson2022bacterial,dunkel2013fluid} and cytoskeletal filaments \cite{needleman2017active, duclos2020topological} to Quincke rollers \cite{zhang2021active, reyes2023magnetic, zhang2023spontaneous}, phenomena such as spontaneous flow, swarming, and dynamic pattern formation emerge. To understand the physics of these systems, several computational models have been developed to simulate their dynamics. Continuum hydrodynamic models represent the orientational order of an active fluid using the  $Q$ tensor field typically used to describe nematic liquid crystalline order \cite{olmsted1992isotropic,toth2002hydrodynamics}. These models are able to determine the flow under various boundary conditions using finite element methods \cite{varghese2020confinement,luo2024flow}. On the other hand, particle-based models, such as Vicsek-style models \cite{vicsek1995novel,chate2020dry}, use unit vectors to represent the direction of self-propulsion of active fluid particles and incorporate a mechanism for aligning these particles. This approach has led to models in which cytoskeletal filaments are represented as semiflexible chains of self-propelled disks \cite{athani2024symmetry, moore2020collective, montesi2005brownian} and to models of perception-dependent active particles \cite{saavedra2024swirling}.

In many practical situations, active fluids interact with passive inclusions, which leads to significant modifications in transport properties. In a passive fluid, the diffusion of the inclusions can only be tuned by the system temperature of the external field directly interacting with the inclusions, such as in the case of DNA polymer in microfluidic chamber \cite{cohen2005method, lameh2020controlled}. In contrast, the diffusivity of the inclusions in active fluid is directly affected by the surrounding active fluid particles. For example, the presence of passive tracers in bacterial suspensions has been shown to exhibit enhanced diffusion \cite{wu2000particle}, and chiral gear-shaped inclusion particles display rectified motion in a bacteria bath \cite{di2010bacterial, sokolov2010swimming} and also an active nematic fluid bath\cite{ray2023rectified}. These inclusion particles create complicated boundary conditions for the active fluid bath, posing great challenges for the continuum models in capturing the localized interactions between the active particles and complex-shaped inclusions. This limitation motivates us to develop a particle-based model that simulates the motion of both the active fluid bath and the inclusion particles.

In this work, we present a particle-based model that simulates the motion of both the self-propelled active particles along with the passive inclusion particles of various shapes using Brownian dynamics simulations \cite{van1982algorithms, huber2019brownian}. The active fluid bath is modeled by a collection of self-propelled, disk-shaped particles that interact through a soft-repulsion interaction in addition to a nematic interaction that favors the orientational alignment of nearby particles. The inclusion particles are modeled as rigid bodies composed of passive disks connected to form predefined rigid shapes, where each disk interacts with an active fluid disk through the soft-repulsion interaction. We consider four shapes for the inclusion particles as shown in Fig.~\ref{fig:inclusion_illustration}: a rod-like $I$ shape, a boomerang-like $L$ shape of two rod-like line segment\cite{chakrabarty2013brownian} and two stair-like $Z$ and $Z^*$ chiral shapes with opposite handedness and three line segments each. The approximate diffusion tensor for each inclusion particle is calculated by considering the contribution from each line segment of the particle while neglecting the interaction between segments. We first investigate the state of the active fluid bath, where a transition between the isotropic and nematic states is determined by the number density of the active fluid particles, the nematic interaction strength and the P\'eclet number. Then, we study the rotational and translational dynamics of the inclusion particles in the active fluid bath, where the interaction with surrounding active fluid particles significantly enhances both the translational and rotational diffusivity of the inclusion particles. Furthermore, chiral inclusion particles exhibit rectified motion, resulting in spontaneous rotation in one direction, while particles of opposite handedness rotate in the opposite sense. Finally, we examine the coupling between the translational and orientational motion of the inclusion particles by measuring the distribution of the translational motion over all orientations relative to the particles' body frame, and we study the dependence of the correlation on the active fluid particle swimming speed.

The rest of this paper is organized as follows. In Section~\ref{sec:method}, we define our particle-based model and explain the Brownian dynamics method we use for the simulations. We present the results of our simulations in Section~\ref{sec:results}. Finally, we summarize our paper in Section~\ref{sec:summary}.

\section{Method}
\label{sec:method}

\subsection{Active fluid bath}
We model the active fluid particles as circular disks on a two dimensional plane. Each disk $i$ is decorated with a unit vector pointing in the swimming direction $\vb{n}_i = (\cos\theta_i,\sin\theta_i)$. The disk-disk interaction $U(i,j)$ has two parts, $U(i,j) = U_r(i,j) + U_\theta(i,j)$:
\begin{equation}
    U(i,j) = \begin{cases}
        \epsilon e^{-(r_{ij}/\sigma)^8} + \frac{K}{2}\cos\theta_{ij} & r_{ij}\leq \sqrt{2}\sigma \\
        0   & r_{ij}  > \sqrt{2}\sigma
    \end{cases}
\end{equation}
where $U_r(i,j) = \epsilon e^{-(r_{ij}/\sigma)^8}$ is a soft repulsive  generalized exponential potential (GEM-8) \cite{moore2020collective}  with cutoff $\sqrt{2}\sigma$ and $U_\theta(i,j)=\frac{K}{2}\cos(\theta_{ij})$ is a nematic interaction that aligns the swimming directions of neighboring beads, with the same cut-off range. 
Here $\sigma$ is the diameter of the disk, $r_{ij}$ is the distance between beads $i$ and $j$, $\theta_{ij}$ is the angle difference between the swimming direction of beads $i$ and $j$, and $\epsilon$ and $K$ are the interaction strengths. 

\subsection{Inclusion particles}
\begin{figure}[!ht]
    \centering
    \includegraphics[width=\linewidth]{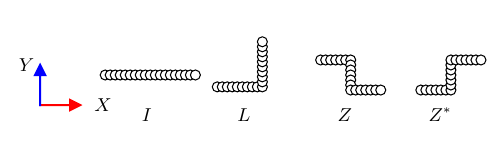}
    \caption{Illustration of the four inclusion particles considered in this work. X and Y are body frame coordinates. Each inclusion particle is made of a total of $M=19$ disks, each with diameter $\sigma$ and the distance between connected disks is $\sigma/2$. The $I$ shape is rod-like, with all disks forming one line segment. The $L$ shape is boomerang-like, such that the disks form two line segments with angle of $\pi/2$ between them. The $Z$ and $Z^*$ shapes are both stair-like with opposite chiral handedness, and the constituent disks form three line segments.}
    \label{fig:inclusion_illustration}
\end{figure}

We model the inclusion particles as rigid bodies of passive disks glued together. Active and inclusion particle interact only via the $U_r$ interaction. The forces and torque exerted on the inclusion particles are calculated based on the summation of the forces exerted by all constituent disks and the corresponding torque with respect to the center of the mass. We consider four kinds of inclusion particles: straight rod, rod with one $\pi/2$ turn (boomerang), and two chiral shapes that have two $\pi/2$ turn but differ in handedness. The inclusion particles are made of $M=19$ disks with diameter $\sigma$, and the distance between connected disks is $\sigma/2$. Fig.~\ref{fig:inclusion_illustration} shows the four types of inclusion particles considered.

It is difficult to precisely calculate the diffusion tensor of these inclusion particles when fully accounting for the complicated hydrodynamic interaction between the particles and the solvent. A precise calculation is beyond the scope of our current study and instead we approximate the diffusivity of these inclusion particles by considering the linear combination of each rod-shape segment, ignoring the hydrodynamic interaction between these line segments, such that the diffusivity in the body frame defined in Fig.~\ref{fig:inclusion_illustration} can be given by $\vb{D}^{b} = diag(D_{\parallel}, D_{\perp})$. For the disk we have the Stokes--Einstein relation $D_t = k_B T/ (3\pi \eta \sigma)$, where $\eta$ is the viscosity, while for a rod or the $I$ particle $\vb{D}^{b}$  is given by \cite{de1981hydrodynamic, lowen1994brownian, de1981hydrodynamic, yang2017interfacial}:
\begin{equation}
\begin{aligned}
    D_\parallel(\gamma) & = \frac{3D_t}{2\gamma} \left[ \ln\left(\gamma\right) - 0.207 + 0.980 \frac{1}{\gamma} -0.1333\frac{1}{\gamma^2} \right] \\
    D_\perp(\gamma) &= \frac{3D_t }{4\gamma} \left[ \ln\left(\gamma\right) + 0.839 + 0.185\frac{1}{\gamma} + 0.2333\frac{1}{\gamma^2} \right] \\
    D_r(\gamma) &= \frac{9D_t}{\sigma^2\gamma^3} \left[ \ln\left(\gamma\right) - 0.662 + 0.917\frac{1}{\gamma} - 0.05\frac{1}{\gamma^2} \right]
\end{aligned}
\end{equation}
where $\gamma = l/\sigma$ is the aspect ratio of the rod with length $l$ and width $\sigma$, $D_\parallel$ is the diffusivity along the long axis, $D_\perp$ is the diffusivity along the short axis and $D_r$ is the rotational diffusivity. $D_t$ and $D_\theta$ are respectively the translational and rotational diffusivity of the disk of diameter $\sigma$. The diffusivities $D_{XX}$, $D_{YY}$ and $D_r$ for each inclusion particle can then be derived and are shown in Table~\ref{tab:diffusion_tensor}. Detailed calculation of these diffusion tensors can be found in Appendix~\ref{sec:diffusion_tensor_for_the_inclusion_particles}

\begin{table}[!ht]
    \centering
    \begin{tabular}{|c|c|c|c|}
    \hline
    shape & $D_{XX}/D_t$ & $D_{YY}/D_t$ & $D_{r}\sigma^2/D_t$ \\ \hline
       $I$ & 0.328838 & 0.237181 & 0.00779303\\  \hline
       $L$ & 0.209881 & 0.209881& 0.0155996 \\ \hline
       $Z,Z^*$ &0.17947 & 0.168594 & 0.0137002 \\ \hline    
    \end{tabular}
    \caption{Diffusion tensor of each inclusion particle in the body frame.}
    \label{tab:diffusion_tensor}
\end{table}

The diffusion tensor in the lab frame can be calculated using:
\begin{equation}
\begin{aligned}
D_{xx} &= D_{XX}\cos^2\Theta + D_{YY}\sin^2\Theta\\
D_{xy} &= (D_{XX}-D_{YY})\cos\Theta \sin\Theta\\
D_{yx} &= (D_{XX}-D_{YY})\cos\Theta \sin\Theta\\
D_{yy} &= D_{XX}\sin^2\Theta + D_{YY}\cos^2\Theta
\end{aligned}
\label{eq:D transformation}
\end{equation}
where $\Theta$ is the orientation of the inclusion particle in the lab frame, and $D_r$ is the same before and after the transformation. 

\subsection{Brownian dynamics simulation}
In a Brownian dynamics simulation scheme \cite{fernandes2002brownian}, we consider the state of a particle represented by a generalized coordinate $\vb{q}(t) = (q_x(t), q_y(t), q_\theta(t))$ acted on by a generalized force
$\vb{F}^{(g)}(t)= (F_x(t), F_y(t), \tau(t))$. The update of the particle's state over a time step $\Delta t$ is given by:
\begin{equation}
    \vb{q}(t+\Delta t) = \vb{q}(t) + \frac{\Delta t}{k_B T}\,\vb{D}\cdot \vb{F^{(g)}}(t) + \delta \vb{q} 
\end{equation}
where \(\vb{D}\) is the generalized diffusion tensor and \(\delta \vb{q}\) is a random increment with covariance $\langle \delta \vb{q}\,\delta \vb{q}^T \rangle = 2\,\vb{D}\,\Delta t$, $k_B$ is the Boltzmann constant and $T$ is the system temperature. Partitioning $\vb{D}$ into its translational and rotational parts, the explicit update equations for $\Delta \vb{q} = \vb{q}(t+\Delta t) - \vb{q}(t)$ become:
\begin{equation}
    \begin{aligned}
    \Delta q_x &= \frac{\Delta t}{k_B T}\Bigl[D_{xx}\,F_x(t) + D_{xy}\,F_y(t)\Bigr] + \delta q_x, \\
    \Delta q_y &= \frac{\Delta t}{k_B T}\Bigl[D_{xy}\,F_x(t) + D_{yy}\,F_y(t)\Bigr] + \delta q_y, \\
    \Delta q_\theta &= \frac{\Delta t}{k_B T}\,D_r\,\tau(t) + \delta q_\theta,        
    \end{aligned}
    \label{equ:Brownian dynamics}
\end{equation}
with the stochastic increments satisfying. $\langle (\delta q_x)^2 \rangle = 2\,D_{xx}\,\Delta t$, $\langle (\delta q_y)^2 \rangle = 2\,D_{yy}\,\Delta t$ $\langle \delta q_x\,\delta q_y \rangle = 2\,D_{xy}\,\Delta t$ and $\langle (\delta \theta)^2 \rangle = 2\,D_r\,\Delta t$.  These update equations apply to both the active fluid and inclusion particles. 

For an individual active fluid particle $i$ at position $(x_i, y_i)$ with swimming direction $\vb{n}_i = (\cos\theta_i,\sin\theta_i)$, the particle state is $\vb{q}_i = (x_i, y_i, \theta_i)$, and the translational force $\vb{F}_i$ and torque $\tau_i$ are given respectively by:
\begin{equation}
    \begin{aligned}
        \vb{F}_i &= v \vu{n}_i - \sum_{j\neq i} \grad U_r(i,j) - \sum_{\alpha} \grad U_r(i,\alpha) \\
        \tau_i &= - \sum_{j\neq i}\grad U_\theta(i,j)
    \end{aligned}
\end{equation}
where $v$ is the swimming speed, $\sum_{j\neq i}$ is a summation over all active fluid disks $j$ other than $i$ and $\sum_{\alpha}$ is the summation over all $M$ passive disks making up an inclusion particle. The active fluid particles' diffusivities $D_{xx}=D_{yy}=D_t$, $D_{xy}=D_{yx}=0$ and swimming speed $v$ are the same for all disks, and the rotational diffusivity is given by $D_r = D_\theta = v/Pe$ where $Pe$ is the P\'eclet number. 

For the inclusion particle, the center of mass is at $(x_c, y_c) = \vb{r}_c$ such that $\vb{r}_c = \sum_\alpha \vb{r}_\alpha/M$, the orientation is given by the angle between body frame and lab frame $\Theta = \arccos{\vb{X}\cdot \vb{x}}$, the particle state is $\vb{q}_p = (x_c, y_c, \Theta)$, the total force and torque act on its center of mass are given by, the
\begin{equation}
    \begin{aligned}
        \vb{F}_p &= - \sum_{i,\alpha} \grad U_r(i,\alpha) \\
        \tau_p &= - \sum_{i,\alpha}(\vb{r}_\alpha - \vb{r}_c)\cross \grad U_r(i,\alpha)
    \end{aligned}
\end{equation}
where $\vb{r}_\alpha$ is the position of disk $\alpha$ on the inclusion particle. The translational and rotational diffusivity of the inclusion particle are given in Table~\ref{tab:diffusion_tensor} and Eq.~\eqref{eq:D transformation}.

To update the state of both the active fluid disks and the inclusion particle, we substitute the appropriate generalized coordinate $\vb{q}$ and force $\vb{F}^{(g)}$ into Eq.~\eqref{equ:Brownian dynamics}. For instance, for an active fluid disk $i$, we let $\vb{q} = (x_i, y_i, \theta_i)$ and $\vb{F}^{(g)} = (\vb{F}_i, \tau_i)$, while for the inclusion particle, we set $\vb{q} = (x_c, y_c, \Theta)$ and $\vb{F}^{(g)} = (\vb{F}_p, \tau_p)$. Finally, we use a mid-step algorithm \cite{fixman1978simulation,hinch1994brownian} to implement the system update during the simulation.


\section{Results}
\label{sec:results}

We first study the active fluid bath system in the absence of inclusion particles, and identify the driving parameters for the states of the active fluid bath. We then add the inclusion particles, which are driven by both their inherent diffusion and the interaction with the active fluid bath. Both the translational and orientational dynamics of the inclusion particles are considered. Finally, we investigate the coupling between the translational motion and the orientation of the inclusion particles in the active fluid bath. In all of our simulations, we use $\epsilon = 20$ for the soft interaction, $\Delta t = 10^{-4}$ for the update time step and set the translational diffusivity of the active fluid particle $D_t=1$ and disk size $\sigma=1$ as natural units. Our simulation box is of size $B\times B$ with periodic boundary condition and we simulate systems with $N=300$ active fluid particles. For the simulations without inclusion particles, we run 10 independent simulations for each set of system parameters for a maximum time $t=100$. For the simulations with inclusion particles, 20 independent simulations are carried out for each system parameters up to $t=200$.

\subsection{States of the active fluid bath}
\begin{figure*}[!t]
    \centering
    \includegraphics{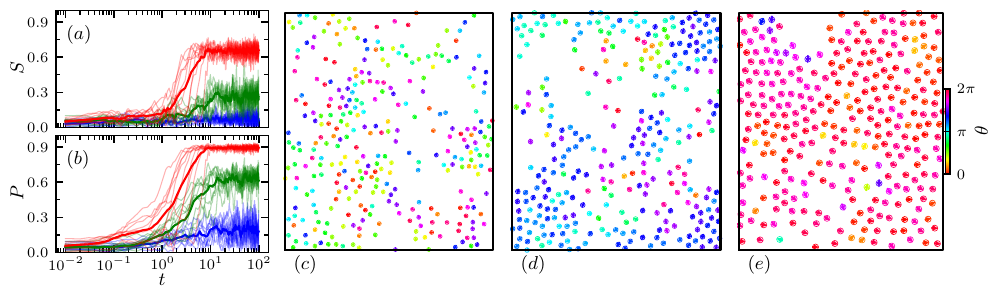}
    \caption{Time evolution of the active fluid with number of disks $N=300$, swimming speed $v=5$, P\'eclet number $Pe=5$, for various values of  density $\rho$ and interaction strength $K$. (a) Nematic order $S$ for $(\rho,K)=(0.2,20)$ (bottom blue line), $(\rho,K)=(0.3,30)$ (middle green line) and $(\rho,K)=(0.5,50)$ (top red line). The light lines are measurements from independent runs, and the bold line is the ensemble average. (b) Similar to (a), but for the polar order $P$. (c) Sample snapshot of the active fluid with $(\rho,K)=(0.2,20)$. (d) Snapshot for $(\rho,K)=(0.3,30)$. (e) Snapshot for $(\rho,K)=(0.5,50)$. The colorbar shows the orientation of the disks.}
    \label{fig:time_evolution}
\end{figure*}
We use both nematic $S=|\left< e^{i 2\theta_i} \right>_i|$ and polar $P=|\left< e^{i\theta_i} \right>_i|$ order parameters to quantify the state of the active fluid, where $\theta_i$ corresponds to the swimming orientation of the disk $i$ relative to the $x$ axis, and  $\left< \dots\right>_i$ is an average over all disks. Figs.~\ref{fig:time_evolution}(a) and (b) show the time evolution of the nematic and polar order respectively for pure active fluid for three different values of the number density $\rho=N/B^2$ and interaction strength $K$. The system is initialized with random orientations of the disks corresponding to $S(0)=P(0)=0$ at time zero. In Figs.~\ref{fig:time_evolution} (a) and (b), the nematic order $S$ and polar order $P$ very quickly increase from zero and reach plateaus. At low $\rho$ and $K$, the disks remain isotropic, as shown in Fig.~\ref{fig:time_evolution}(c), leading to low values of both $S$ and $P$. As both density $\rho$ and interaction strength $K$ increase, the system transitions to the nematic state, as shown in Fig.~\ref{fig:time_evolution}(e), quantified by large nematic order $S$ and polar order $P$.

\begin{figure}[!htb]
    \centering
    \includegraphics{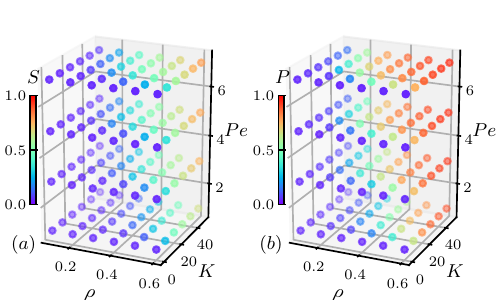}
    \caption{Phase diagram of the active fluid system with number of disks $N=300$, quantified by nematic and polar order versus density $\rho$, interaction strength $K$ and P\'eclet number $Pe$. (a) Nematic order $S=|\left< e^{i 2\theta_i} \right>_i|$. (b) Polar order $P=|\left< e^{i\theta_i} \right>_i|$.}
    \label{fig:phase_diagram}
\end{figure}

The state of the active fluid system is also affected by the P\'eclet number $Pe = v/D_\theta$, as higher $Pe$ corresponds to more ballistic motion of these particles. Fig.~\ref{fig:phase_diagram} shows the time averaged nematic and polar order for various values of the density $\rho$, interaction strength $K$ and P\'eclet number $Pe$. The nematic polar order is averaged over times $t>25$, where the system has entered equilibrium, indicated by the time evolution shown in Figs.~\ref{fig:time_evolution} (a) and (b). Both nematic order $S$ and polar order $P$ increase with increasing density $\rho$, interaction strength $K$, and P\'eclet number $Pe$. The increased density reduces the relative distance between disks, enhancing the effect of the aligning interaction. Increasing $K$ induces stronger torque which also tends to align the disks, and finally, the higher P\'eclet number $Pe$ effectively lowers the noise level for the orientational diffusion, making the system more orientationally ordered.

\begin{figure}[!htb]
    \centering
    \includegraphics{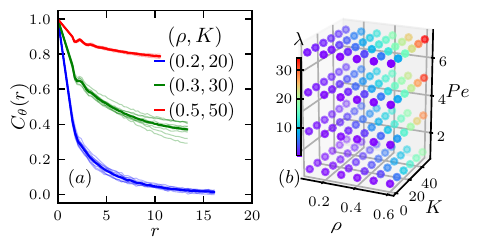}
    \caption{Orientation correlation of the active fluid bath. (a) Orientational correlation function $C_\theta(r)$ for various combination of density $\rho$ and interaction strength $K$. (b) Correlation length $\lambda$ versus density $\rho$, interaction strength $K$ and P\'eclet number $Pe$.}
    \label{fig:correlation_function}
\end{figure}

In addition to the nematic and polar order, we also measure the orientational correlation function of the active fluid, which captures the decay of relative orientation alignment as the distance between two pair of disks increases. The orientational correlation is defined as 
\begin{equation}
    C_\theta(r) =  \frac{\sum_{i,j} \cos(\theta_i-\theta_j) \delta( |\vb{r}_i - \vb{r}_j| - r)} {\sum_{i,j}\delta( |\vb{r}_i - \vb{r}_j| - r)}
\end{equation}
where $\delta$ is the Kronecker delta function, $r$ is the distance between two disks, and the sum $\sum_{i,k}$ is the over all pair of disks including self-pairing. As shown in Fig.~\ref{fig:correlation_function} (a) the correlation function $C_\theta(r)$ exhibits exponential decay, and we define the correlation length $\lambda$ by fitting the correlation function using $C_\theta(r) = e^{-r/\lambda}$. Fig.~\ref{fig:correlation_function} (b) shows the correlation length averaged over 10 independent runs for various values of density $\rho$, interaction strength $K$ and P\'eclet number, reiterating the results discussed above about the state of the active fluid as the correlation length increases when the system transitions from the isotropic to nematic phase.

\subsection{Dynamics of the inclusion particles}
\begin{figure}[!htb]
    \centering
    \includegraphics{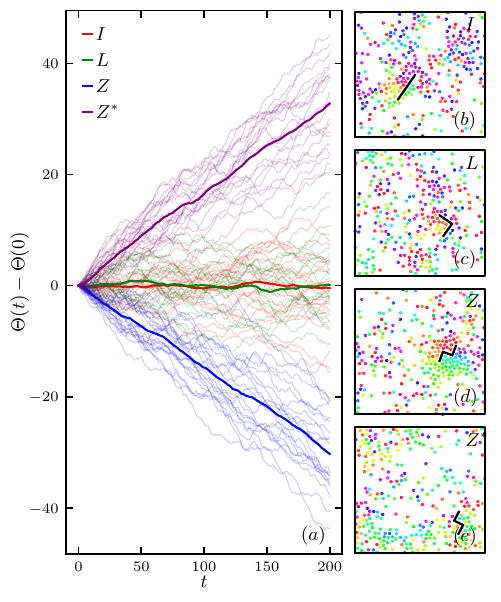}
    \caption{Rotational dynamics of four types of inclusion particles with number of active fluid particles $N=300$, density $\rho=0.2$, swimming speed $v=5$, interaction strength $K=10$, P\'eclet number $Pe=5$ (a) Time evolution of inclusion particle orientation. There are 20 independent runs for each inclusion particle, denoted by the light lines, and the bold line represents the average over all runs. (b)-(e) Snapshot of the inclusion particles in the active fluid bath for $I$, $L$, $Z$ and $Z^*$ particles, respectively. The color represents the orientation of the active fluid disks using the same color bar as in Fig.~\ref{fig:time_evolution}.}
    \label{fig:inclusion_type}
\end{figure}

Now we consider the addition of inclusion particles of various shapes to the active fluid bath and study the rotational and translational dynamics of the inclusion particles. We consider four types of inclusion particles: $I$-shape, $L$-shape, left-handed $Z$-shape and right-handed $Z$-shape (see Fig.~\ref{fig:inclusion_illustration}).  Each of these particles consists of 19 connected disks of the same size as the active fluid particles. The distance between connected disks is $0.5$ and each arm of the particles has the same length. Interacting with the surrounding active fluid, the inclusion particles rotate in a stochastic manner. While the $I$ and $L$ particles' orientation evolve in Brownian motion with near zero mean value, as shown in Fig.~\ref{fig:inclusion_type}(a), the chiral inclusion particles, including both the $Z$ and $Z^*$, exhibit a constant drift corresponding to a single rotational direction. Moreover, the left-handed $Z$ and right-handed $Z^*$ rotate in opposite directions.

\begin{figure}[!htb]
    \centering
    \includegraphics{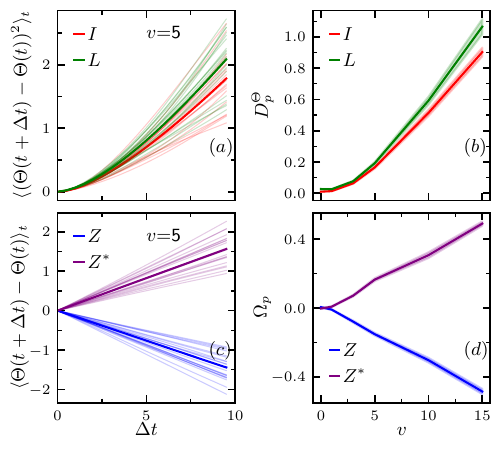}
    \caption{Rotational diffusivity and  speed for different inclusion particles with number of active fluid particles $N=300$, density $\rho=0.2$, swimming speed $v=5$, interaction strength $K=10$ and P\'eclet number $Pe=5$. (a) Mean square displacement (MSD) of the orientation $\Theta$ for the $I$ and $L$ particles. (b) Inclusion particle rotational diffusivity $D^\Theta$ fitted from the MSD versus active fluid bath swimming speed $v$.  (c) Mean displacement of the end-to-end orientation $\theta$ for the chiral $Z$ and $Z^*$ particles. (d) Rotational velocity $\Omega_p$ for chiral particles versus active fluid bath swimming speed $v$.}
    \label{fig:inclusion_diffusivity_rotation}
\end{figure}

To quantify the rotational diffusivity and rotational speed of these inclusion particles, we calculate the mean square displacement (MSD) of orientation $\Theta$ for the $I$ and $L$ particles, and  the orientational mean displacement for the chiral $Z$ and $Z^*$ particles. Fig.~\ref{fig:inclusion_diffusivity_rotation}(a) shows the MSD $\left<(\theta(t+\Delta t) - \theta(t))^2 \right>_t$ versus time interval $\Delta t$, which exhibits a linear trend for $\Delta t\gtrsim2$, where $\left<\dots\right>_t$ denotes the average over time. Fitting the slope of the MSD yields the corresponding rotational diffusivity $D^\Theta$ for $I$ and $L$ shape particles. The diffusivity is much greater than the inherent Brownian diffusivity shown in Table~\ref{tab:diffusion_tensor}, indicating it is driven by the interaction with the surrounding active fluid bath. Fig.~\ref{fig:inclusion_diffusivity_rotation} (b) shows that the rotational diffusivity $D^\Theta$ increases almost linearly with the swimming speed of the active fluid particles. As for the chiral particles, a constant shift speed is observed in Fig.~\ref{fig:inclusion_type} (a), which is due to the symmetry breaking of the torque applied by the surrounding active fluid bath. Fig.~\ref{fig:inclusion_diffusivity_rotation} (c) shows that the mean displacement $\left<\theta(t+\Delta t) - \theta(t) \right>_t$ is non-zero, and increases linearly with the time interval $\Delta t$ for the left-handed $Z$ particle, and decreases linearly for the right-handed $Z^*$ particle. Correspondingly, Fig.~\ref{fig:inclusion_diffusivity_rotation} (d) shows the rotational speed $\Omega_p$, fitted from the slope of the mean displacement versus time interval, also increases/decreases linearly with the swimming speed of active fluid particles.

\begin{figure}[!htb]
    \centering
    \includegraphics{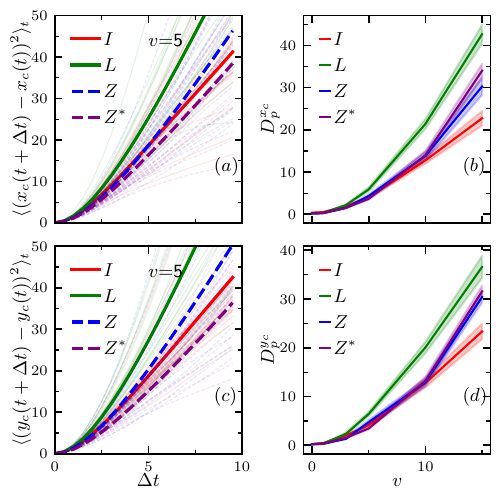}
    \caption{Translational diffusivity and speed for different inclusion particles with number of active fluid particles $N=300$, density $\rho=0.2$, swimming speed $v=5$, interaction strength $K=20$ and P\'eclet number $Pe=5$. (a) Mean square displacement of the center of mass in the $x$ direction of the lab frame $x_c$ for all four types of particles. (b) Diffusivity of the inclusion particles in the $x$ direction $D^{x_c}_p$ versus active fluid bath swimming speed $v$. (c) Similar to (a) but for the $y$ direction $y_c$. (d) Similar to (b) but for the $y$ direction.}
    \label{fig:inclusion_diffusivity_translation}
\end{figure}

Moreover, the translational diffusivity is also subject to the interaction with the active fluid bath. Fig.~\ref{fig:inclusion_diffusivity_translation} (a) shows the MSD of the particle center of the mass in the $x$ direction of the lab frame $\left<(x_c(t+\Delta t) - x_c(t))^2 \right>_t$, which exhibits a similar trend as the orientation $\theta$. Similarly, fitting the slope gives the translational diffusivity along the $x$ direction $D^{x_c}_p$. As shown in Fig.~\ref{fig:inclusion_diffusivity_translation} (b), $D^{x_c}_p$ increases with increasing swimming speed $v$ of the active fluid bath. Figs.~\ref{fig:inclusion_diffusivity_translation} (c) and (d) show similar results for the $y$ direction in the lab frame.

\subsection{Coupling between translation and orientation}
\begin{figure}[!htb]
    \centering
    \includegraphics{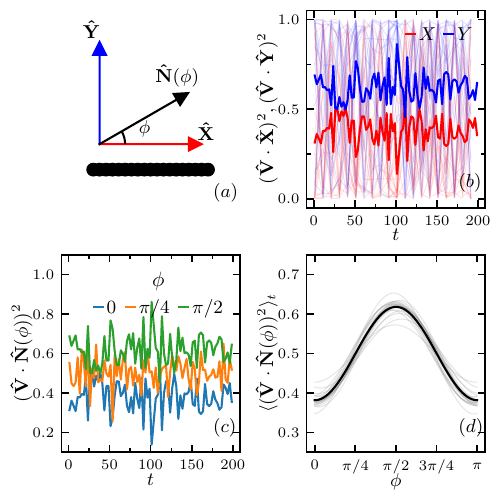}
    \caption{Coupling between the translational motion and the orientation of the inclusion particle with number of active fluid particles $N=300$, density $\rho=0.2$, swimming speed $v=15$, interaction strength $K=20$ and P\'eclet number $Pe=5$ for the $I$ particle . (a) Illustration of the reference direction $\vu{N}(\phi)=\vu{X}\cos\phi+\vu{Y}\sin\phi$ in the body frame ($X,Y$). (b) Time series of the particle translational motion $\vu{V}=\vb{V}/|\vb{V}|$, projected to the $\vu{X}$ and $\vu{Y}$ direction of the body frame, light line are different run, and bold line is the average over all 20 runs. (c) Generalized coupling between translational motion $\vu{V}$ and the reference direction $\vu{N}(\phi)$, $\phi=0,\pi/2$ corresponds to $\vu{X}$ and $\vu{Y}$ directions, respectively. (d) Time averaged coupling $\vu{V}\cdot\vu{N}(\phi)$ versus $\phi$.}
    \label{fig:I_diffusion_coupling_process}
\end{figure}

We also investigate the coupling between the translational motion of the inclusion particles and their orientation. Because the paths of Brownian motion of the particles are not continuous and thus not differentiable, we define the translational velocity of the particles using the finite difference
\begin{equation}
    \vb{V}(t)=\left[\vb{r}_c(t+\Delta_m t/2)-\vb{r}_c(t-\Delta_m t)\right]/\Delta_m t   
\end{equation}
where $\vb{r}_c = (x_c, y_c)$ is the center of mass of the particle in the lab frame, and we use $\Delta_m t=0.5$ for our calculation. In order to examine the coupling between the translational velocity $\vb{V}$ and the particle orientation, we define a reference unit vector $\vu{N}(\phi) = \vu{X}\cos\phi+\vu{Y}\sin\phi$ using the $X$ and $Y$ components of the unit vector in the body frame as illustrated in Fig.~\ref{fig:I_diffusion_coupling_process}(a). By projecting the unit length translational velocity $\vu{V}=\vb{V}/|\vb{V}|$ onto the reference unit vector $\vu{N}(\phi)$, we can investigate the overall distribution of translational motion relative to the body frame. Fig.~\ref{fig:I_diffusion_coupling_process}(b) shows the coupling between $\vu{V}$ and the body frame velocity components $\vu{X}$ and $\vu{Y}$ when the active fluid bath swimming speed $v=15$. The average from different runs shows clear discrepancy between the $X$ direction $(\vu{V}\cdot\vu{X})^2$ and $Y$ direction $(\vu{V}\cdot\vu{Y})^2$, with the motion along the $Y$ direction favored over the $X$ direction. This is contrary to the implication of the inherent Brownian diffusivity of the $I$ shape particle in the body frame as shown in Table~\ref{tab:diffusion_tensor}, in which the $I$ shape particle diffuses faster along the long axis $X$ as compared to along the short axis $Y$. Again, this is due to the interaction with surrounding active fluid bath, such that the longer interaction length along the $Y$ direction provide the $I$ shape particle more exposure to the interaction force from the active fluid particles. Moving on to include more directions, Fig.~\ref{fig:I_diffusion_coupling_process}(c) shows the coupling between $\vu{V}$ and three different directions of the reference unit vector $\vu{N}(\phi)$. By averaging the coupling $(\vu{V}\cdot\vu{N}(\phi))^2$ over time, and adding more directions $\phi$, Fig.~\ref{fig:I_diffusion_coupling_process}(d) shows the coupling $\left<(\vu{V}\cdot\vu{N}(\phi))^2\right>_t$ for different reference directions, and illustrates that the active fluid bath drives the $I$ shape particle to move more along the short axis than the long axis.

\begin{figure}[!ht]
    \centering
    \includegraphics{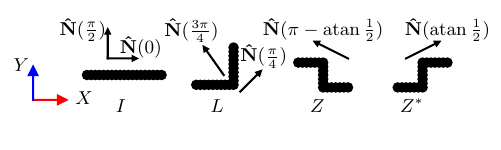}
    \caption{Characteristic orientation in the body frame for different inclusion types.}
    \label{fig:characteristic_angle}
\end{figure}

While it is natural to quantify the characteristic direction of the $I$ shape particle using $\vu{X}$ and $\vu{Y}$, it is not obvious for the  particles with other shapes. In Fig.~\ref{fig:characteristic_angle}, we attempt to illustrate the characteristic direction of each type of particles. For the $L$ shape, we annotate the end-to-end direction $\vu{N}(\frac{\pi}{4})$ as well as the perpendicular direction. For the $Z$ and $Z^*$ particles, we annotate the end-to-end directions as shown.

\begin{figure}[!ht]
    \centering
    \includegraphics[width=\linewidth]{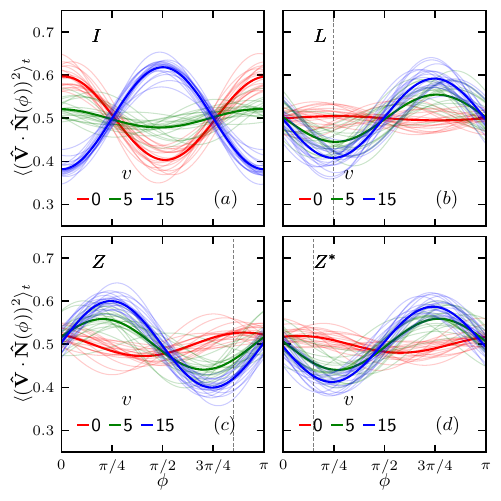}
    \caption{Correlation between translational velocity $\vu{V}=\vb{V}/|\vb{V}|$ and orientation $\vu{N}=\vu{X}\cos\phi+\vu{Y}\sin\phi$ for inclusion particles of different shapes with number of active fluid particles $N=300$, density $\rho=0.20$, interaction strength $K=20$ and P\'eclet number $Pe=5$ for various active fluid bath swimming speed $v$. Light lines are different independent runs and bold lines are averages over these runs. (a)-(d) are results for $I$, $L$, $Z$ and $Z^*$ particles, respectively. For $L$, $Z$ and $Z^*$, the dashed lines correspond to orientations $\phi=\pi/4$, $\pi-\atan\frac{1}{2}$ and $\atan\frac{1}{2}$, respectively }
    \label{fig:inclusion_diffusion_correlation}
\end{figure}

We then carry out an analysis similar to the one shown in Fig.~\ref{fig:I_diffusion_coupling_process}(d) for three different values of swimming speed $v$ and all four types of inclusion particles. Fig.~\ref{fig:inclusion_diffusion_correlation} (a) shows the translation-orientation coupling $\left<(\vu{V}\cdot\vu{N}(\phi))^2\right>_t$ of the $I$ shape particle for active fluid particles swimming speed $v=0,5$ and $15$. When $v=0$ , the active fluid bath is passive, and the Brownian motion of the inclusion particle $I$ is purely driven by its inherent diffusivity. This is reflected by the higher coupling at $\phi=0$ and $\pi$ corresponding to the long axis versus the short axis $\phi=\pi/2$. As the swimming speed $v$ increases, the correlation becomes higher along the short axis and lower along the long axis as shown in Fig.~\ref{fig:I_diffusion_coupling_process}(d). For the $L$ shape particle, when the swimming speed $v=0$, the correlation $\left<(\vu{V}\cdot\vu{N}(\phi))^2\right>_t$ is nearly independent of $\phi$,  as $D_{XX} = D_{YY}$ for the $L$ shape particle. Increasing the swimming speed $v$ (i.e. activity of the bath) reduces the correlation  along the end-to-end direction $\phi=\frac{\pi}{4})$ and increases the correlation along $\phi=\frac{3\pi}{4}$, as indicated by the dashed line in Fig.~\ref{fig:inclusion_diffusion_correlation}. This is because when the active fluid particles bump into an $L$ shape particle along the $\vu{N}(\frac{\pi}{4})$ direction, they are colliding from the side, and can easily exert non-zero net torque to rotate the $L$ shape particle, whereas when the active fluid particles are colliding from the $\vu{N}(\frac{3\pi}{4})$ direction, the $L$ shape particle can remain stable. For the chiral $Z$ particles, Fig.~\ref{fig:inclusion_diffusion_correlation}(c) shows the particle is more prone to move along the end-to-end direction when $v=0$ for the active fluid bath, and the correlation $\left<(\vu{V}\cdot\vu{N}(\phi))^2\right>_t$ is reversed as the swimming speed $v$ of active fluid particle increases. A $Z^*$ shape particle shows a trend similar to $Z$ as shown in Fig.~\ref{fig:inclusion_diffusion_correlation}(d).

\section{Summary}
\label{sec:summary}
In this work, we studied the dynamics of rigid inclusion particles immersed in an active fluid bath using Brownian dynamics simulations. Our study explored the states of the active fluid bath and focused on the interplay between the active fluid bath and the shapes of the inclusion particles. The active fluid is modeled as a collection of self-propelled circular disks that interact through a soft repulsive potential and an nematic interaction that encourages the alignment between nearby disks. The properties of the active fluid bath are tuned by the swimming speed $v$, density $\rho$, nematic interaction strength $K$ and P\'eclet number $Pe$. The inclusion particles are modeled as rigid assemblies of disks. Four types of inclusion particles were constructed, denoted as $I$, $L$, $Z$ and $Z^*$. Their inherent diffusivity was approximated by considering the contributions from each rod-like segments. The $I$ and $L$ shape particles are achiral, while the $Z$ and $Z^*$ shape particles are chiral and of opposite handedness. 

Our simulations show that the active fluid transitions from an isotropic to a nematic phase with increasing density, interaction strength and P\'eclet number. This transition is characterized by the nematic order, polar order and orientational correlation length. When inclusion particles are added, the active fluid bath drives the inclusion particles to diffuse more strongly than their inherent diffusivity. In addition, the chiral inclusion particles $Z$ and $Z^*$ exhibit nonzero average angular speed with opposite directions due to the opposite handedness. The translational motion of these inclusion particles is also coupled with their orientation, and the correlation is tuned by the swimming speed of the active fluid bath.

Given the large number of parameters in the system, we have only studied a limited subset and have only investigated the active fluid bath with a single inclusion particle. Future research should explore other system parameters, for example, the effect of the swimming speed on the states of the active bath as well as the effects of the active fluid density, interaction strength and P\'eclet number on the dynamics of the inclusion particles. Moreover, our framework is general enough to be expanded to study inclusion particles with arbitrary shapes, for example, $L$ shape and $Z$ shape particles with arms of different length. Finally, when adding more than one inclusion particle, the interaction between these particles can be studied, especially the impact of handedness on the interaction between chiral inclusion particles.

\section{Data Availability}
The code for the simulation and data analysis are available at the GitHub repository \href{https://github.com/ljding94/active_fluid_2d}{active fluid 2d}

\section{Author Contributions}
LD, RAP and TRP conceived the work; LD derived the theoretical framework, developed the simulation code, generated and analyzed the simulation data; and LD, RAP, and TRP wrote and edited the manuscript.

\section{Conflicts of interest}
There are no conflicts to declare.

\begin{acknowledgments}
We thank Daniel Beller for fruitful discussions. This work was supported by the National Science Foundation through Grant No. MRSEC DMR-2011846 and CBET-2227361. This research was performed at the Spallation Neutron Source, which is DOE Office of Science User Facilities operated by Oak Ridge National Laboratory. This research was sponsored by the Laboratory Directed Research and Development Program of Oak Ridge National Laboratory, managed by UT-Battelle, LLC, for the U.S. Department of Energy. This research used resources of the Oak Ridge Leadership Computing Facility, which is supported by the DOE Office of Science under Contract DE-AC05-00OR22725. This research used resources of the National Energy Research Scientific Computing Center, which is supported by the Office of Science of the U.S. Department of Energy under Contract No. DE-AC02-05CH11231. 
\end{acknowledgments}

\bibliography{reference}
\clearpage

\onecolumngrid
\appendix

\section{Diffusion tensor of the inclusion particles}
\label{sec:diffusion_tensor_for_the_inclusion_particles}

The diffusivity in the body frame is given by $\vb{D}^{b} = diag(D_{\parallel}, D_{\perp})$. Note that for the disk,  $D_t = k_B T/ (3\pi \eta \sigma)$, where $\eta$ is the viscosity. The $\vb{D}^{b}$ for the $I$ particle is given by \cite{de1981hydrodynamic, lowen1994brownian, de1981hydrodynamic, yang2017interfacial}
\begin{equation*}
\begin{aligned}
    D_\parallel(\gamma) & = \frac{3D_t}{2\gamma} \left[ \ln\left(\gamma\right) - 0.207 + 0.980 \frac{1}{\gamma} -0.1333\frac{1}{\gamma^2} \right] \\
    D_\perp(\gamma) &= \frac{3D_t }{4\gamma} \left[ \ln\left(\gamma\right) + 0.839 + 0.185\frac{1}{\gamma} + 0.2333\frac{1}{\gamma^2} \right] \\
    D_r(\gamma) &= \frac{9D_t}{\sigma^2\gamma^3} \left[ \ln\left(\gamma\right) - 0.662 + 0.917\frac{1}{\gamma} - 0.05\frac{1}{\gamma^2} \right]
\end{aligned}
\end{equation*}
where $\gamma = l/\sigma$, $l$ is the length of the rod and $\sigma$ is the width of the rod. 

\subsection{Translational diffusion}
For the translational motion, we use the simple linear combination of different rod-shape segments in each inclusion particle. We note that the translational friction $\zeta_\parallel \equiv k_B T/D_\parallel$ and $\zeta_\perp = k_B T/D_\perp$. All of the rod-shape segments in these inclusion particles are either along the $X$ direction or the $Y$ direction in the body frame. Thus, the off-diagonal term of the diffusion tensors are all zero, such that $D^{L}_{XY}=D^{L}_{YX} = 0$, $D^{Z}_{XY}=D^{Z}_{YX} = 0$, and $D^{Z^*}_{XY}=D^{Z^*}_{YX} = 0$.

For each inclusion particle, the total length is $(M+1)\sigma/2$, denoting the aspect ratio $\Gamma = (M+1)/2$, the diffusion tensor for the $I$ shape rod-like particle is $D^I_{XX} = D_\parallel(\Gamma)$, $D^I_{YY} = D^I_\perp(\Gamma)$, $D^I_{XY} = D{YX}=0$ and $D^I_r = D_r(\Gamma)$.

For the $L$ shape particle, there is one rod with $\Gamma/2$ along $X$ and another rod with $\Gamma/2$ along the $Y$ direction, thus the friction $\zeta_X = \zeta_Y = \zeta_\perp(\Gamma/2) + \zeta_\parallel(\Gamma/2)$, ignore the segment-segment interaction, for $L$ shape
\begin{equation}
    \begin{aligned}
        D^{L}_{XX} &= \left[ \frac{1}{D_\parallel(\Gamma/2)} + \frac{1}{D_\perp(\Gamma/2)}\right]^{-1} \\
        D^{L}_{YY} &= \left[ \frac{1}{D_\parallel(\Gamma/2)} + \frac{1}{D_\perp(\Gamma/2)}\right]^{-1} \\
    \end{aligned}
\end{equation}

Similarly, for the $Z$ and $Z^*$ shape particles, they both have 2 rods with $\Gamma/3$ along the X direction and 1 rod with $\Gamma/3$ along the Y direction, thus for the $Z$ shape:
\begin{equation}
    \begin{aligned}
        D^{Z}_{XX} &= \left[ \frac{2}{D_\parallel(\Gamma/3)} + \frac{1}{D_\perp(\Gamma/3)}\right]^{-1} \\
        D^{Z}_{YY} &= \left[ \frac{1}{D_\parallel(\Gamma/3)} + \frac{2}{D_\perp(\Gamma/3)}\right]^{-1} \\
    \end{aligned}
\end{equation}
and $D^{Z^*} = D^{Z}$.

\subsection{Rotational diffusion}
To derive the rotational diffusion of these inclusion particles, we need to consider the torque contributed by both  the rotation and translation of the constituent rod-shape segments.

\subsubsection{$L$ shape}
When the $L$ shape particle rotates about its center of mass with angular velocity $\omega$, each rod’s center moves in a circle of radius $d$. For the horizontal rod (aligned along $x$), its velocity due to rotation is given by  $\vec{v}_h = \omega\, \hat{z} \times \Delta \vec{r}_h$. Relative to the center of mass, the center of the horizontal rod is at $\vec{r}_h=(-l/8,-l/8)$. Since $\hat{z}\times (x,y) = (-y,x)$, we obtain $\vec{v}_h = \omega(\frac{l}{8},-\frac{l}{8})$.

Thus, in the frame of the horizontal rod the velocity components are:
$$
v_{h,x} = \frac{\omega l}{8} \quad (\text{along the rod, i.e. } x), \qquad
v_{h,y} = -\frac{\omega l}{8} \quad (\text{perpendicular to the rod, i.e. } y)
$$
The corresponding frictional forces are:
$$
F_{h,x} = -\zeta_{\parallel}(\Gamma/2) \cdot \frac{\omega l}{8}, \quad
F_{h,y} = \zeta_{\perp}(\Gamma/2) \cdot \frac{\omega l}{8}.
$$
The torque about the center of mass due to the horizontal rod is then $\tau_h = (\Delta x)\,F_{h,y} - (\Delta y)\,F_{h,x}$, where $\Delta x = -l/8$ and $\Delta y = -l/8$. Hence,
$$
\tau_h = \frac{l}{8}\Bigl(\zeta_{\perp}(\Gamma/2)\frac{\omega l}{8}\Bigr) + \frac{l}{8}\Bigl(\zeta_{\parallel}(\Gamma/2)\frac{\omega l}{8}\Bigr)
= \frac{l^2\,\omega}{64}\Bigl[\zeta_{t,\parallel}(l/2)+\zeta_{t,\perp}(l/2)\Bigr].
$$
Thus for each rod the total extra contribution is 
$$
\Delta \zeta_r^{(\text{rod})} = \frac{l^2}{64}\Bigl[\zeta_{\parallel}(\Gamma/2)+\zeta_{\perp}(\Gamma/2)\Bigr].
$$
A similar calculation for the vertical rod yields the same torque. Taking the rotational friction from both rods into account, $\zeta_r^{(\text{rod})} = k_B T/D_r(\Gamma/2)$, the total rotational friction of the particle is $\zeta^L_r = 2\zeta_r^{(\text{rod})}+2\Delta \zeta_r^{(\text{rod})}$, thus the total rotational diffusivity of the $L$ shape particle is

\begin{equation}
  D_r^{L} = \left\{\frac{2}{D_r(\Gamma/2)} + \frac{\Gamma^2\sigma^2}{32}\Bigl[\frac{1}{D_{\parallel}(\Gamma/2)}+\frac{1}{D_{\perp}(\Gamma/2)}\Bigr]\right\}^{-1}  
\end{equation}

\subsubsection{$Z$ shape}
The $Z$ and $Z^*$ shape particles each have three rod-shape segments, two horizontal and one vertical. For the vertical segment, its center of mass is the same as the whole particle, thus there is only inherent rotational friction contributing to the inclusion particle's rotational diffusivity. For the two horizontal segments, we consider the upper right segments in the $Z^*$ shape particle (others are similar). The center of mass of this horizontal rod is at  $\vec{r}_h=(l/6,l/6)$, leading to $\vec{v}_h = \omega(-\frac{l}{6},\frac{l}{6})$.
Thus, in the frame of the horizontal rod the velocity components are:
$$
v_{h,x} = -\frac{\omega l}{6} \quad (\text{along the rod, i.e. } x), \qquad
v_{h,y} = \frac{\omega l}{6} \quad (\text{perpendicular to the rod, i.e. } y)
$$
The corresponding frictional forces are:
$$
F_{h,x} = \zeta_{\parallel}(\Gamma/3) \cdot \frac{\omega l}{6}, \quad
F_{h,y} = -\zeta_{\perp}(\Gamma/3) \cdot \frac{\omega l}{6}.
$$
The torque about the center of mass due to the horizontal rod is then $\tau_h = (\Delta x)\,F_{h,y} - (\Delta y)\,F_{h,x}$, where $\Delta x = l/6$ and $\Delta y = l/6$. Hence,
$$
\tau_h = \frac{l}{6}\Bigl(\zeta_{\perp}(\Gamma/3)\frac{\omega l}{6}\Bigr) + \frac{l}{6}\Bigl(\zeta_{\parallel}(\Gamma/3)\frac{\omega l}{6}\Bigr)
= \frac{l^2\,\omega}{18}\Bigl[\zeta_{\parallel}(\Gamma/3)+\zeta_{\perp}(\Gamma/3)\Bigr].
$$
Thus, for each horizontal rod the total extra contribution is 
$$
\Delta \zeta_r^{(\text{rod})} = \frac{l^2}{18}\Bigl[\zeta_{\parallel}(\Gamma/3)+\zeta_{\perp}(\Gamma/3)\Bigr].
$$
Similarly, the other horizontal rod segment contribute the same torque, and the total rotational friction is consistent with the rotational friction of each rod, $\zeta_r^{(\text{rod})} = k_B T/D_r(\Gamma/3)$. There is also an extra contribution $\Delta \zeta_r^{(\text{rod})}$, such that $\zeta^L_r = 3\zeta_r^{(\text{rod})}+2\Delta \zeta_r^{(\text{rod})}$. Thus, the total rotational diffusivity of the $Z$ and $Z^*$ shape particles are

\begin{equation}
    D_r^{Z} = \left\{\frac{3}{D_r(\Gamma/3)} + \frac{\Gamma^2\sigma_2}{9}\Bigl[ \frac{1}{D_{\parallel}(\Gamma/3)}+\frac{1}{D_{\perp}(\Gamma/3)}\Bigr] \right\}^{-1} 
\end{equation}

\subsection{Transformation to the lab frame}

Assume that in the body frame the diffusion tensor is
$$
\mathbf{D}' = \begin{pmatrix} D_{XX} & 0 \\ 0 & D_{YY} \end{pmatrix},
$$
and that the body’s $X$–axis makes an angle $\Theta$ with the lab-frame $x$–axis. The rotation matrix is
$$
R = \begin{pmatrix} \cos\Theta & -\sin\Theta \\[0.5em] \sin\Theta & \cos\Theta \end{pmatrix}.
$$
Then the diffusion tensor in the lab frame is given by
$$
\mathbf{D} = R\,\mathbf{D}'\,R^T.
$$
A straightforward calculation yields
\begin{equation}
\centering
\begin{aligned}
D_{xx} &= D_{XX}\cos^2\Theta + D_{YY}\sin^2\Theta \\
D_{xy} &= (D_{XX}-D_{YY})\cos\Theta\,\sin\Theta \\
D_{yx} &= (D_{XX}-D_{YY})\cos\Theta\,\sin\Theta \\
D_{yy} &= D_{XX}\sin^2\Theta + D_{YY}\cos^2\Theta
\end{aligned}
\end{equation}

\end{document}